# Multimedia Storage in the Cloud using Amazon Web Services: Implications for Online Education

Santiago Obrutsky, Emre Erturk

Eastern Institute of Technology, New Zealand

**Abstract**

This report is divided into three parts. Firstly, an explanation of the cloud storage products is given, in particular, the Amazon Simple Storage Service (S3). Secondly, the report discusses a case study about SmugMug's migration to the cloud hosted by Amazon. SmugMug is a premium online photo and video sharing service business, and currently has billions of photos and videos from amateur and professional photographers around the world. This report also includes a step by step explanation of how to modify a website and use Amazon S3 to host videos in the cloud. These findings are evaluated in terms of how useful Amazon S3 can be for an e-learning institute, especially one that is geared towards mobile delivery of content. Finally, the paper covers the copyright, and privacy implications, in order to understand the larger context.

**Introduction**

The Amazon Simple Storage Service (Amazon S3) solution is a cloud based storage solution that is part of Amazon Web Services (AWS) and can provide high levels of scalability and flexibility. The customer pays only for the storage that they use with no minimum fees and no initial costs. (Amazon, 2016a). A good example of a business using Amazon S3 to host multimedia content is SmugMug (Amazon, 2016b). SmugMug is a premium online photo and video sharing service business, which stores billions of photos and videos. Because the company had grown very fast and did not have the required in-house hardware capacity, SmugMug opted for Amazon S3 storage capabilities and continued to adopt additional services such as Amazon Elastic Compute Cloud (EC2) and Amazon Cloud Search. Thus, when a customer uploads photos or videos on SmugMug, it is placed on AWS. According to SmugMug's CEO, it was "the biggest, scariest engineering change we've made in the company's history" (MacAskill, 2013). However, the current website is fast and the upload process is quick.

This paper aso gives an overview of the configuration of Amazon S3. This includes practical instructions and steps to upload a video file, modify security options, and modify a website HTML to host a video in the cloud.

Schools and universities also have information technology needs, and need to make some of the similar decisions that businesses do. Along with the other cloud service providers, Amazon has always been active in the educational sector, providing IT infrastructure services to universities and schools (Yoo, 2014). The particular task that is analysed in this paper is using cloud based storage for educational videos and multimedia, especially for e-learning, hybrid learning, and online schools. Erturk (2013b) discusses a mobile learning institute, including its educational philosophy, its information system, and its extensive use of multimedia. For a newly established and agile organization, outsourcing the physical storage of data to a cloud service provider can be a good IT strategy. Aside from Amazon, Rackspace is another enterprise storage service option (Obrutsky, 2016). When it comes to data privacy regulations, security, and geographic proximity, Amazon has an advantage as it operates more data centres in different parts of the world (Dang, 2014).

**Background: Amazon Simple Storage Service (Amazon S3)**

Amazon currently has the highest market share among the providers of cloud hosting services for businesses (Synergy Research Group, 2016). In particular, there are various types of cloud data storage and solutions that Amazon provides. Object Storage is for applications that require scaling and flexibility. It can also be used to import existing data stores for analytics, backup, or archiving. File Storage, i.e. the Elastic File System (EFS), is for applications that need to access shared files and require a file system. It is also suitable for large content repositories, development environments, media stores, and company home directories. Block Storage, i.e. Elastic Block Store (EBS), is for applications such as large databases and ERP systems that often require dedicated and low latency storage. (Amazon, 2016c). Amazon Simple Storage Service (Amazon S3) is another cloud storage option with high security, reliability and scalability. Amazon S3 has web service interface to store and retrieve the data from the internet.

Within Amazon S3, there is a range of storage classes. Amazon S3 Standard is for general-purpose storage of frequently accessed data. Amazon S3 Standard - Infrequent Access (Standard - IA) is for long-lived, but less frequently accessed data. Amazon S3 Glacier is for long-term data archiving (Amazon, 2016c). Once a company starts using S3, the data can be easily migrated from one storage type to another, in order to ensure that the most appropriate service is provided. Amazon S3 is recommended for content distribution, disaster recovery, digital data, backup and recovery, and for cloud based applications.

**Case Study Example**

SmugMug initially selected AWS in 2006 for Amazon S3 storage space, but then adopted additional services, for example, Amazon EC2 for enabling quicker uploads and providing better performance (Amazon, 2016b). SmugMug did not have the sufficient in-house storage space for billions of photos and videos from customers. Because customers entrusted their photos and videos to be safe in the Cloud, SmugMug also needed to store multiple copies in multiple geographic locations, which was possible using Amazon S3. SmugMug also considered other service providers that could host photos and videos in the cloud. However, MacAskill found out that "the pricing was out of our reach, and it wasn't simple to engineer" (Gainesville Sun, 2008, para. 15).

In early 2006, SmugMug was growing successfully but became unable to cope with the growing demand. With a tight budget and only one developer, the company needed to have digital storage space that was inexpensive, simple and reliable. "We looked at Amazon S3's pricing, design and ease-of-use, and were blown away. Amazon S3 is simple and elegant, so much so that it was basically a drop-in addition to our current infrastructure," said MacAskill. The adoption of Amazon S3 took only five days, and saved the company $500,000 in just four months (Business Wire, 2006, para. 4). Before the migration AWS, SmugMug kept all the data in their own local storage drives. However, AWS offered SmugMug superior performance and reliability. Sometimes, the relations between both companies were tested. According to Thomas (2008), Amazon S3 suffered two outages in 2008. The first was in February for two hours, and the second in July for eight hours. However, SmugMug CEO Don MacAskill mentioned that his faith in AWS was not shaken.

SmugMug has now migrated all of its services from its own data servers and onto the AWS Cloud, and is saving time from managing physical hardware, and is gaining greater operational efficiency (Amazon, 2016b). According to Kelleher (2006), Amazon and SmugMug formed a strategic relationship. In 2006, SmugMug was already saving $400,000 annually by using S3, and started considering other Amazon cloud based services. Some of the database and network capacity benefits that SmugMug receives from AWS are:

- Scalability: AWS can store vast numbers of photos and videos into the cloud.
- Durability: AWS provides almost 100% durability because of the data replication that AWS provides.
- Security: AWS has received many industry-recognized certifications and audits.
- Fast Deployment Time: No necessity to wait for purchasing new hardware.
- High Availability: AWS provides 99.99% availability with multiple fault-tolerant availability zones (Amazon, 2016c).

Then in 2011, SmugMug implemented EC2 to improve the processing performance and lower latency during. the uploading, processing, and rendering of photos and videos. It also implemented Amazon Cloud Search to improve the content search experience for users. This service speeds up the search engine to facilitate people finding galleries, sets of photos, and individual photos, very quickly and precisely (MacAskill, 2013, 00:01:43).

**Practical Tutorial: Using Amazon S3 to host a video.**

An AWS account can be created online at https://aws.amazon.com, and includes free storage up to 5 GB. Secondly, it is recommended that the user installs the S3Fox organizer extension for Mozilla Firefox and set up their AWS account. Thirdly, the following steps show how to create a bucket on AWS and upload a video in mp4 format. Lastly, the website is modified (using simple HTML) in order to publish a video using Amazon (S3).

1. The first step it to create an AWS account. Once the organization has an AWS account and is logged in, they can choose the S3 option in the console.

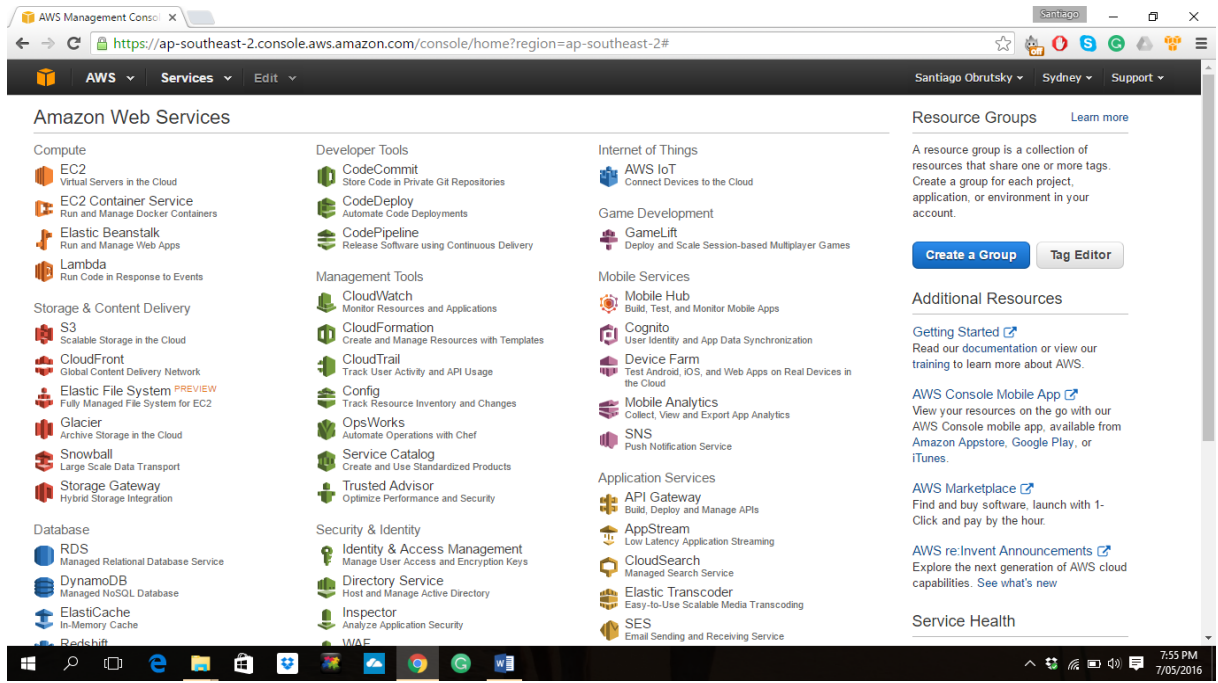

2. In S3, all the buckets that the user creates will be listed on the screen. A bucket is how Amazon names the folders in the cloud.

3. This tutorial uses Mozilla Firefox, therefore the S3Fox organizer extension needs to be installed.

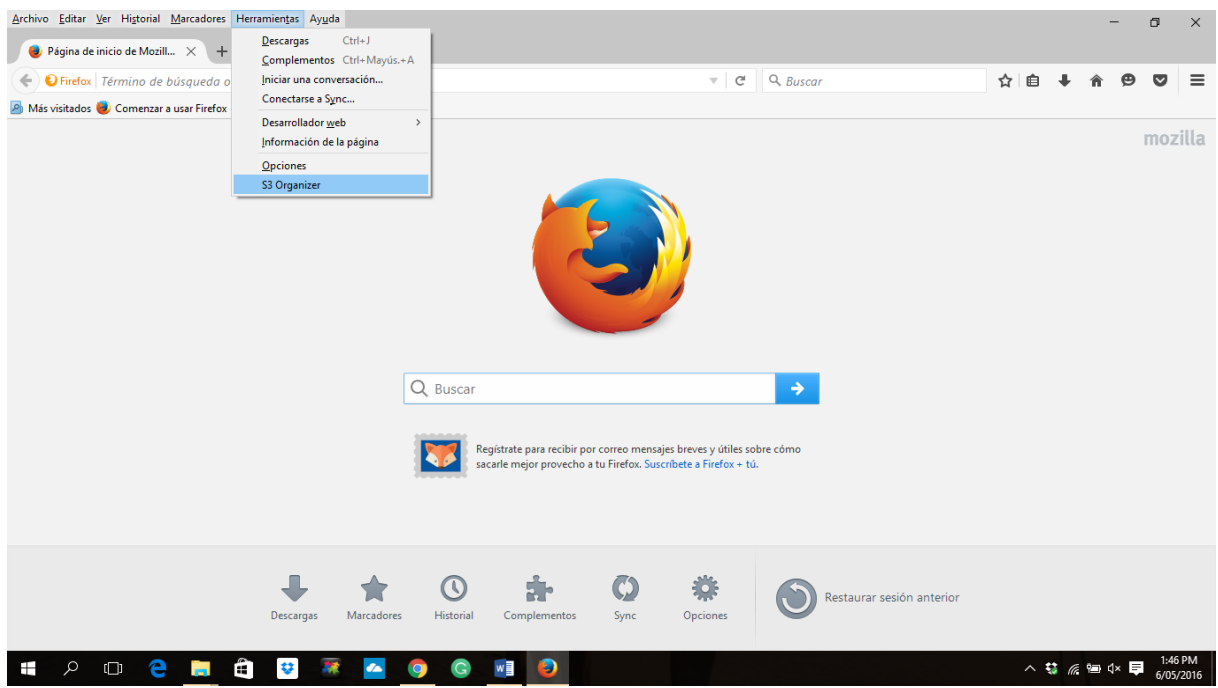

4. The S3Fox Organizer can be found under the Mozilla Add-ons page.

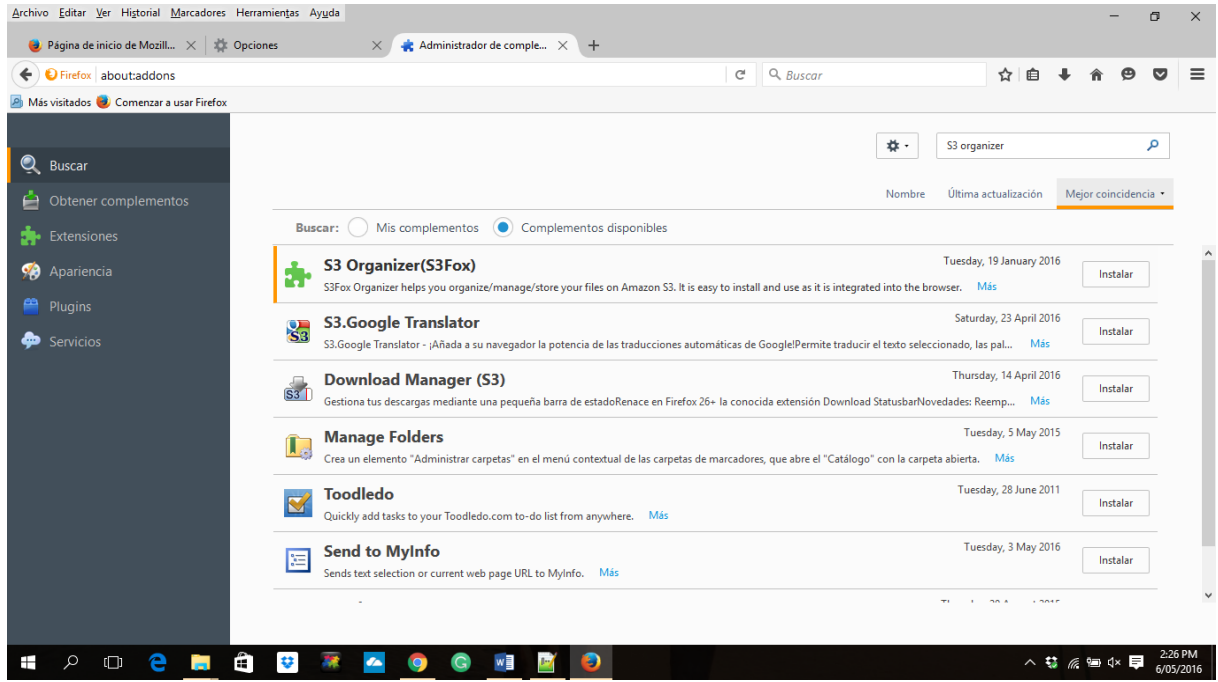

5. The next step is to click on "Manage Accounts" to enter an access key and a code. These two credentials are for the AWS account.

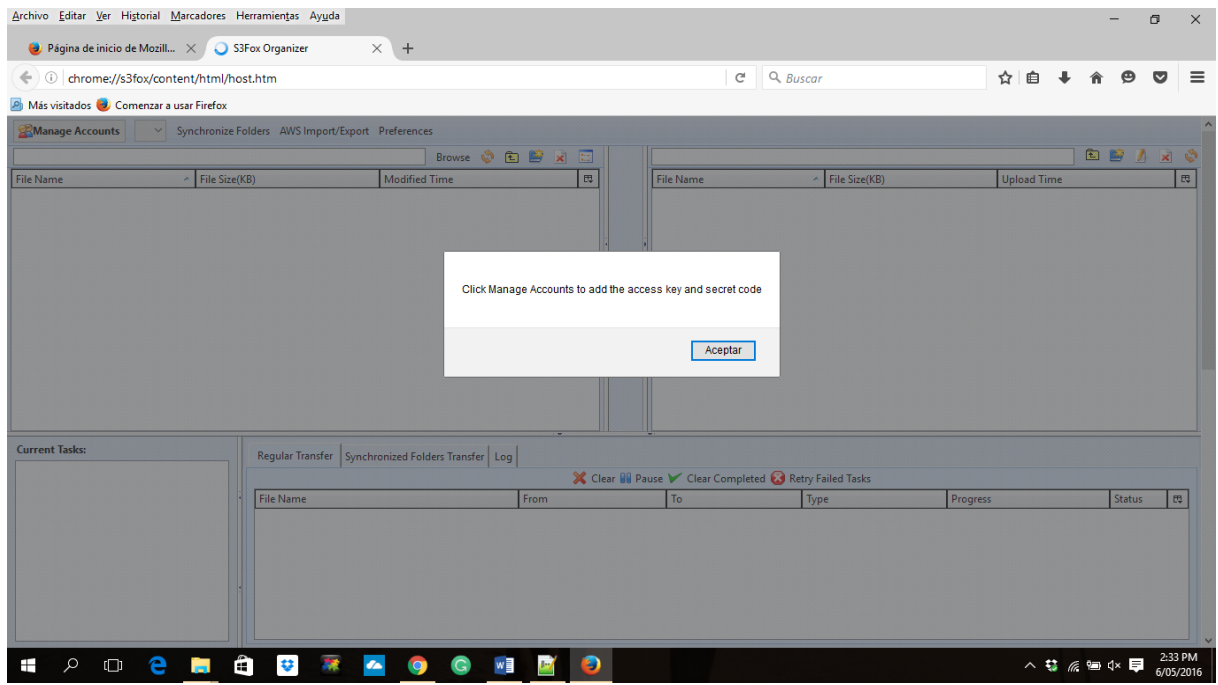

6. To look up the access key and the code, go to the AWS account and select "Security Credentials." Then, select the option "Access Keys" and click on "Create New Access Key."

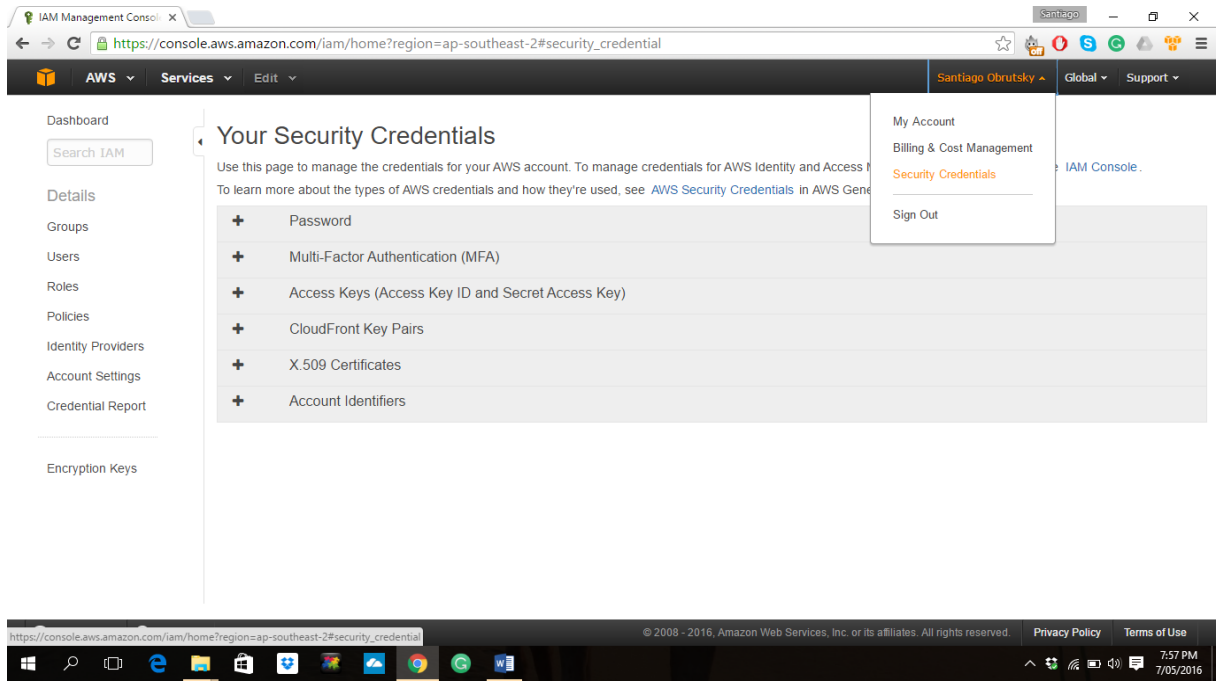

7. The Amazon website will generate a .xls file. The access key and the secret code are in this file.

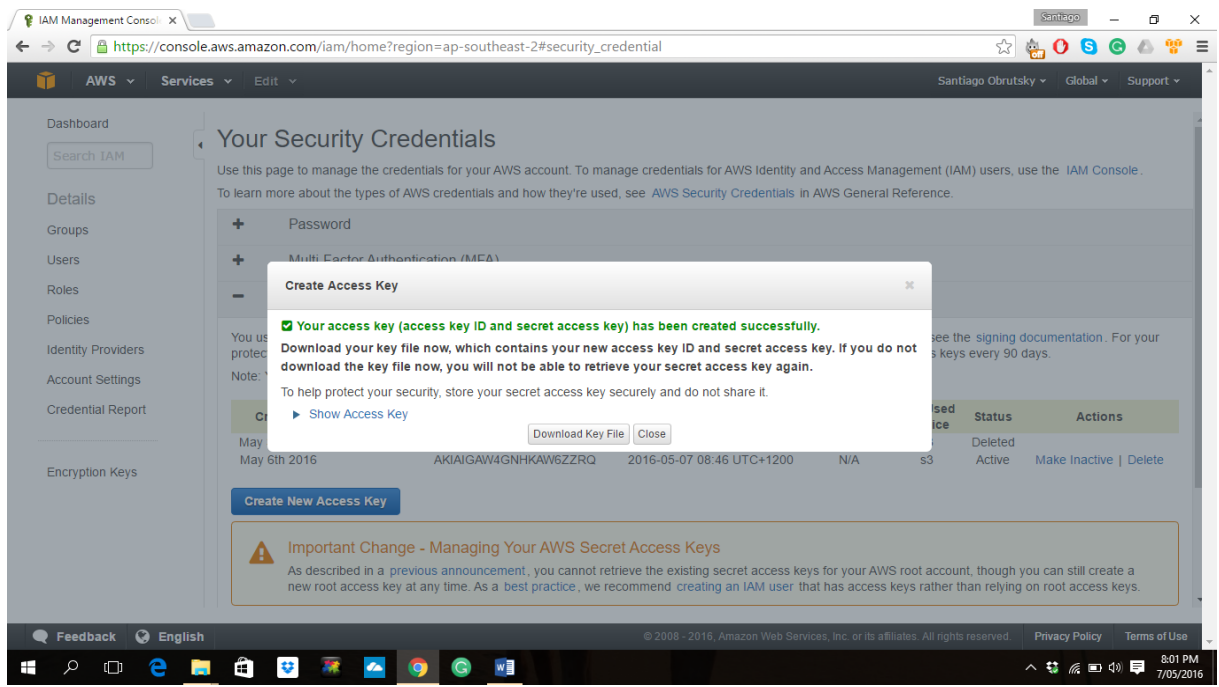

8. In AWS, you can view the running log that shows the access keys generated so far (i.e. the Access Key ID), and their creation dates.

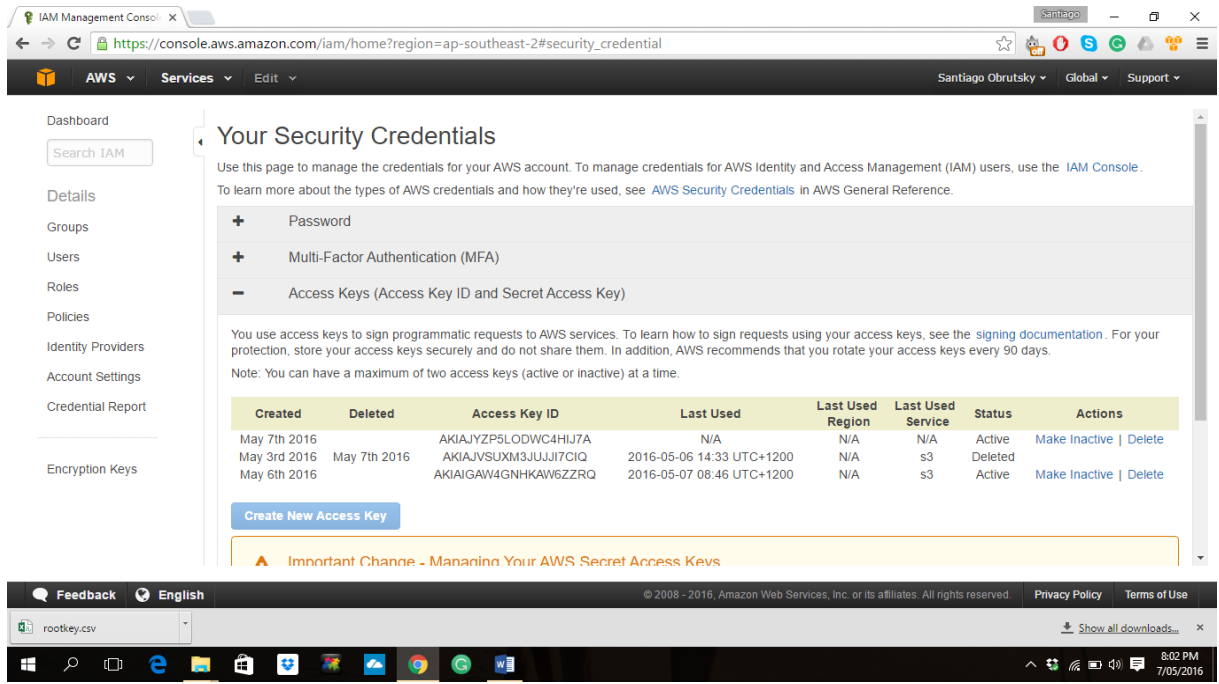

9. Next, go back to the Account Manager in the S3Fox organizer extension. An account is created by using the provided access key and the code.

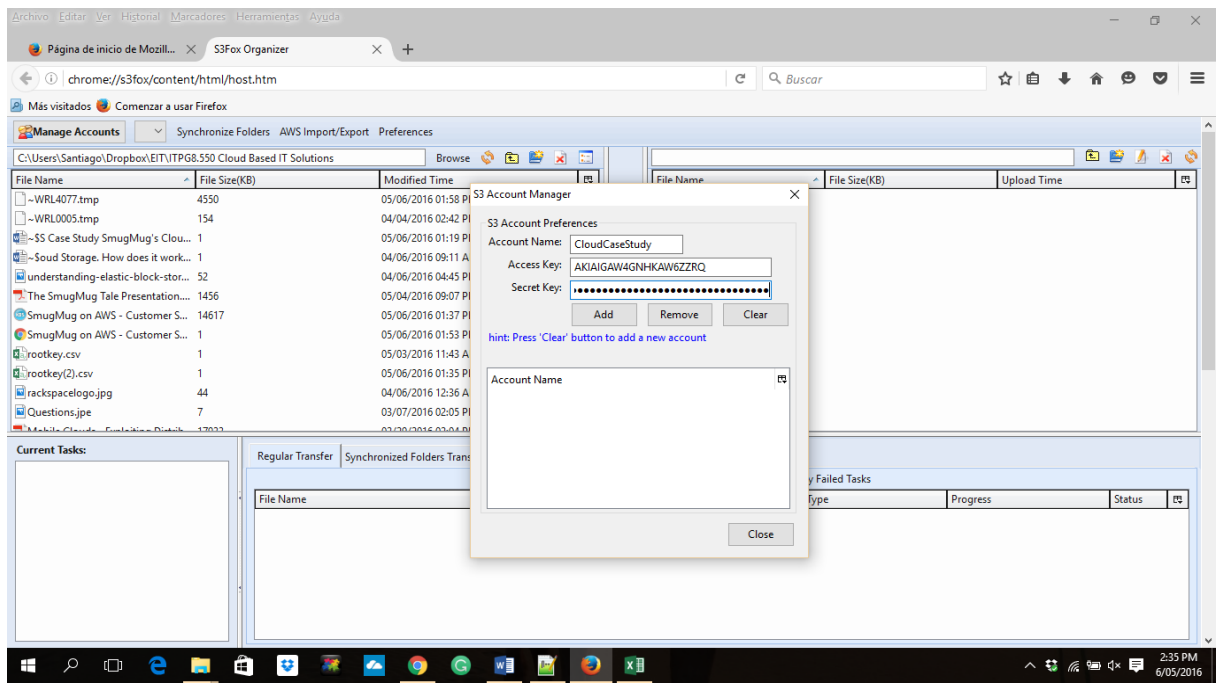

10. After the account creation, the organizer will show the local computer files on the left, and the files uploaded to AWS on the right side.

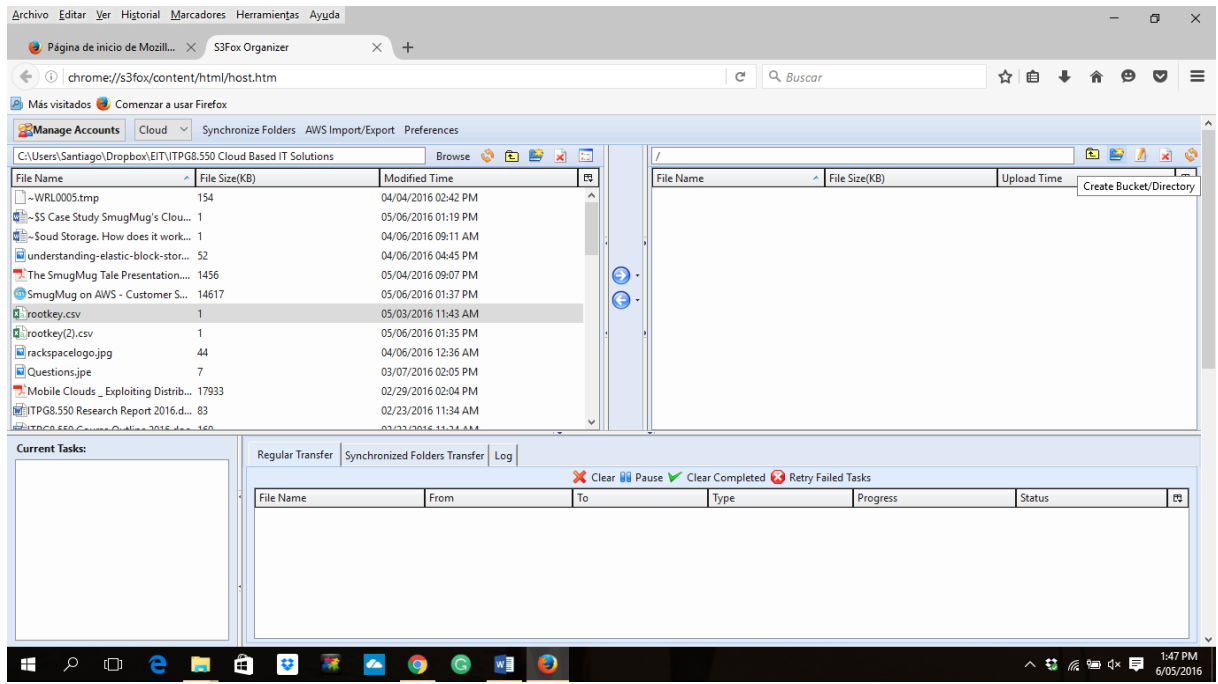

11. The next step is to create a folder, i.e. a bucket. After clicking on "create folder", a pop up will appear for typing the folder name.

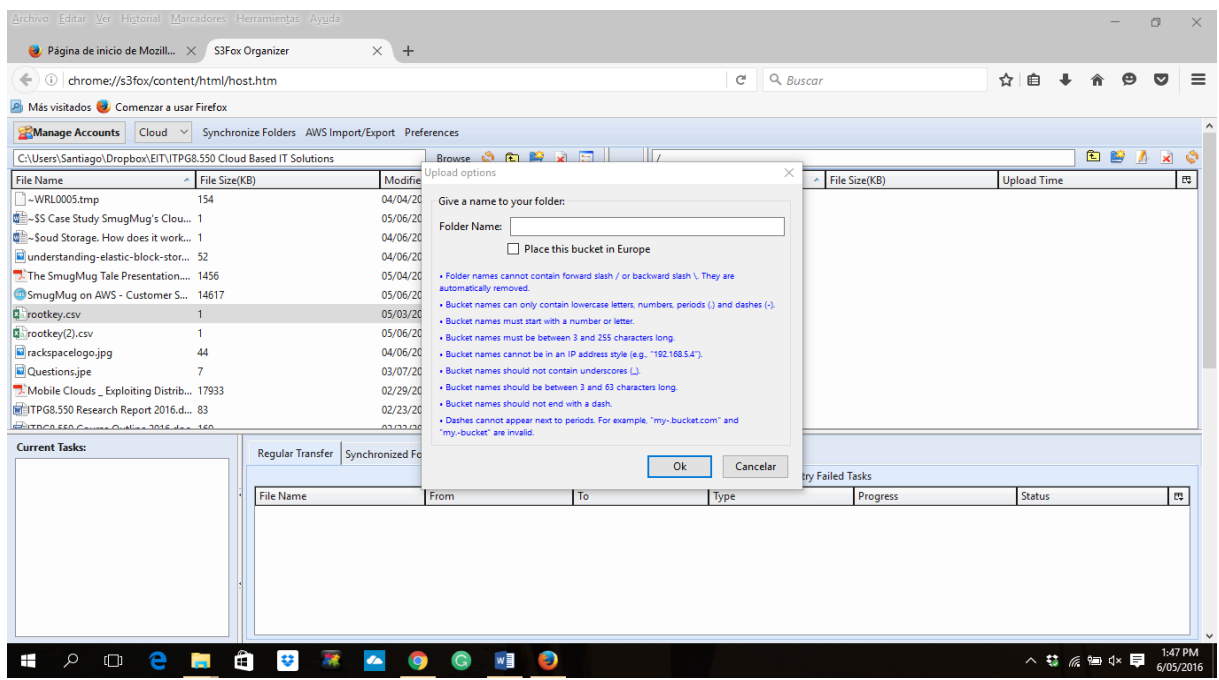

12. After the folder is named, click on "OK" and the bucket will be created on AWS.

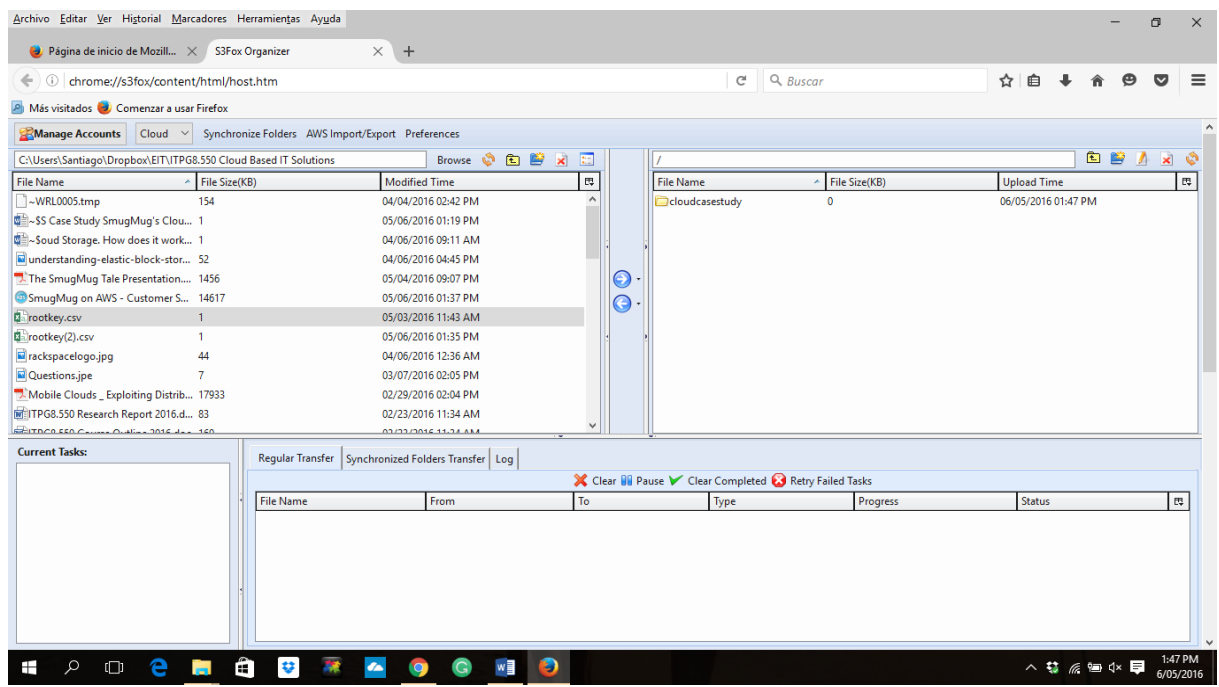

13. After creating the bucket, the first video can be uploaded to Amazon Simple Storage Service (S3). Select the file from the left hand side, and drag and drop into the bucket. This will upload the file to the cloud. Alternatively, it is possible to drag and drop a folder, in order to upload multiple files at the same time.

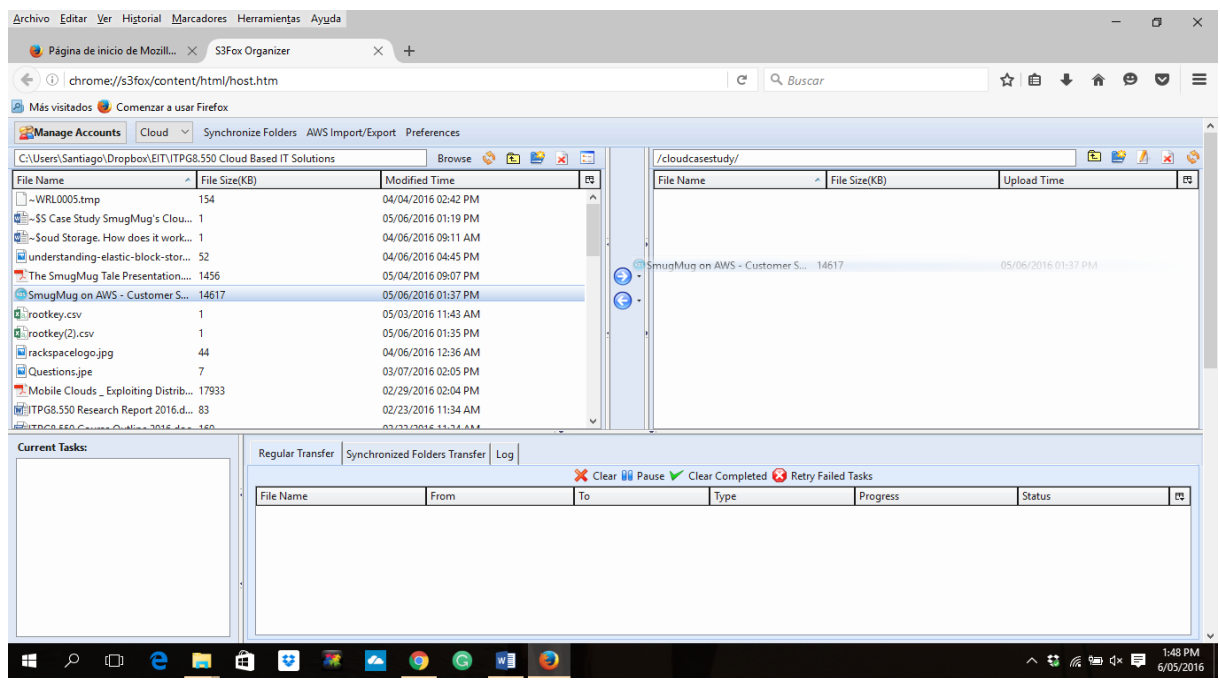

14. After the upload completion, our file will be available for using it from the cloud.

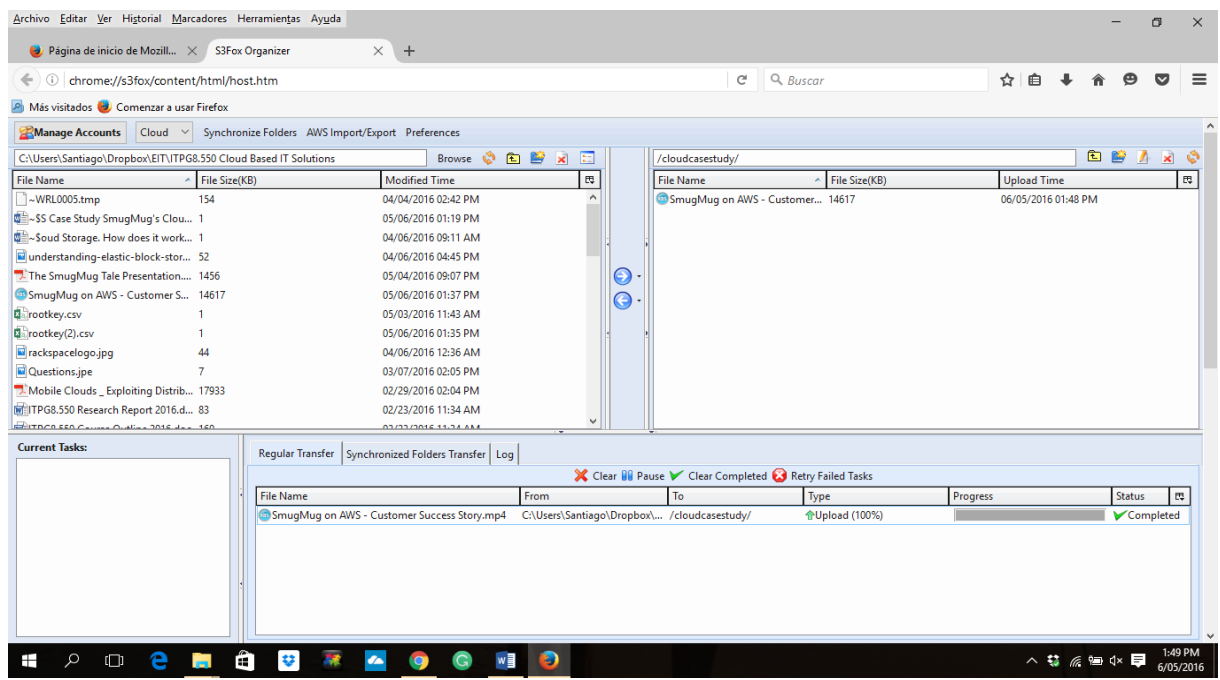

15. The next step is to work on the web page. The following is basic HTML code to provide an example for this tutorial. We can see in the initial HTML code that the video is in the local C: drive in the user's computer.

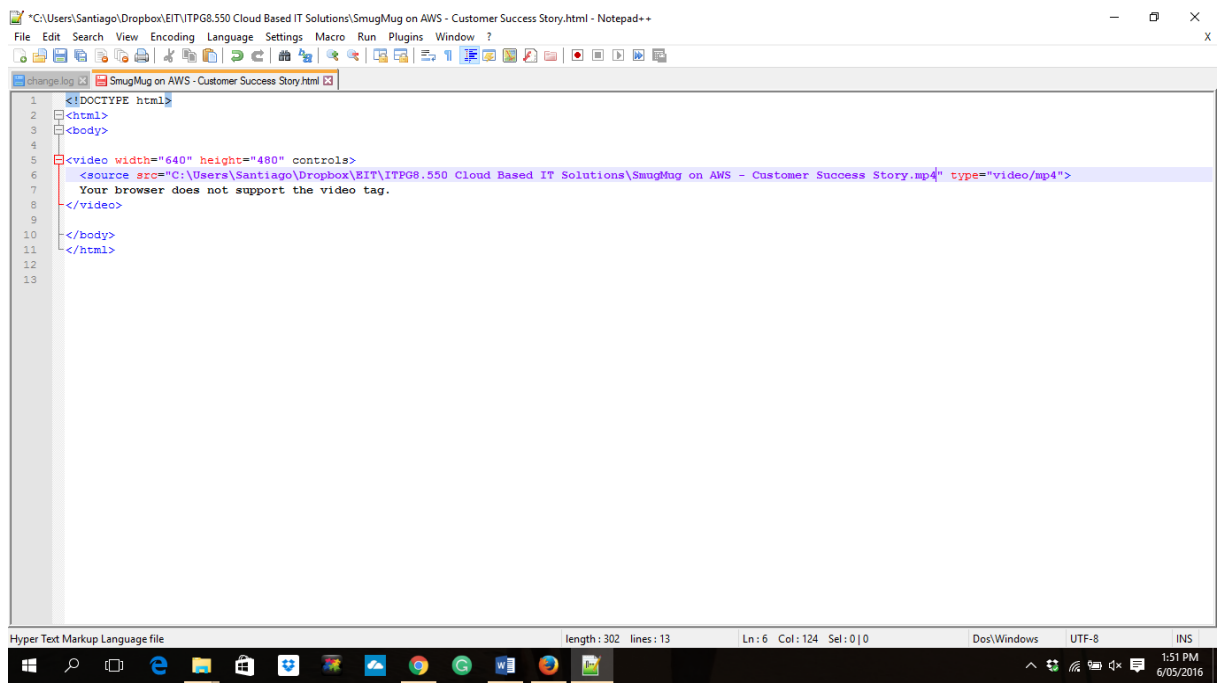

16. Next, go to the S3Fox organizer extension and copy the video URL from AWS by right clicking on it.

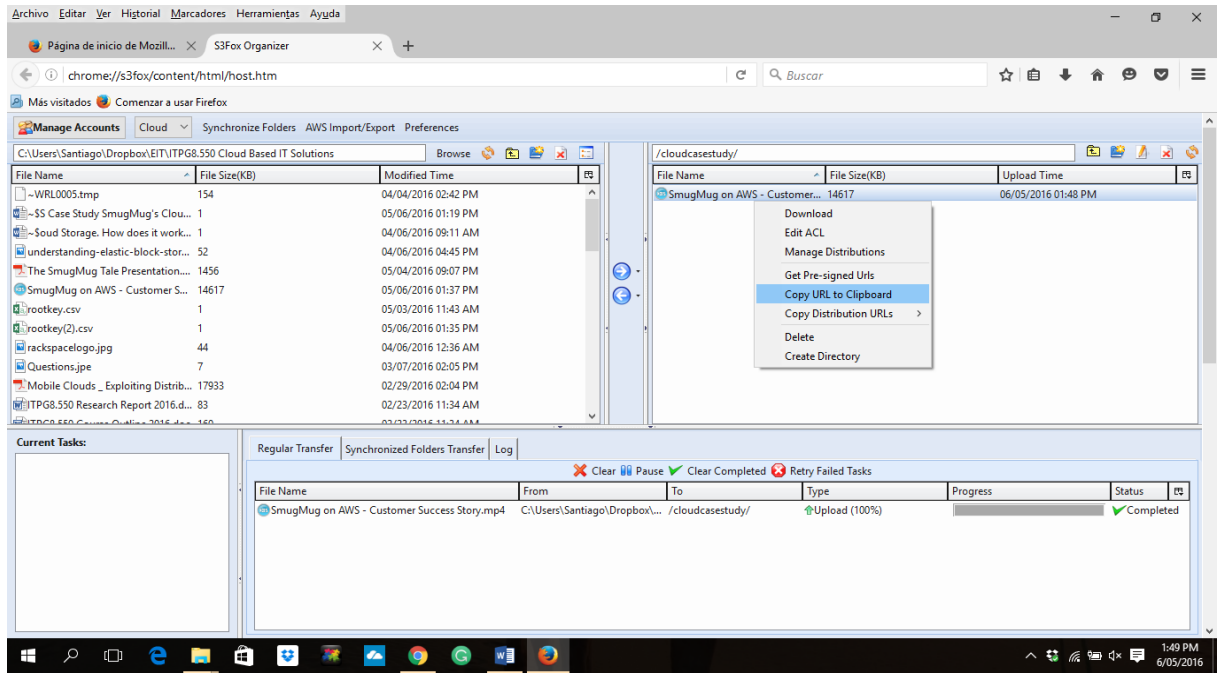

17. Then, go to the HTML file and modify the code. AWS generates a URL automatically in the following format: http://bucketname.s3.amazonaws.com/filename.ext

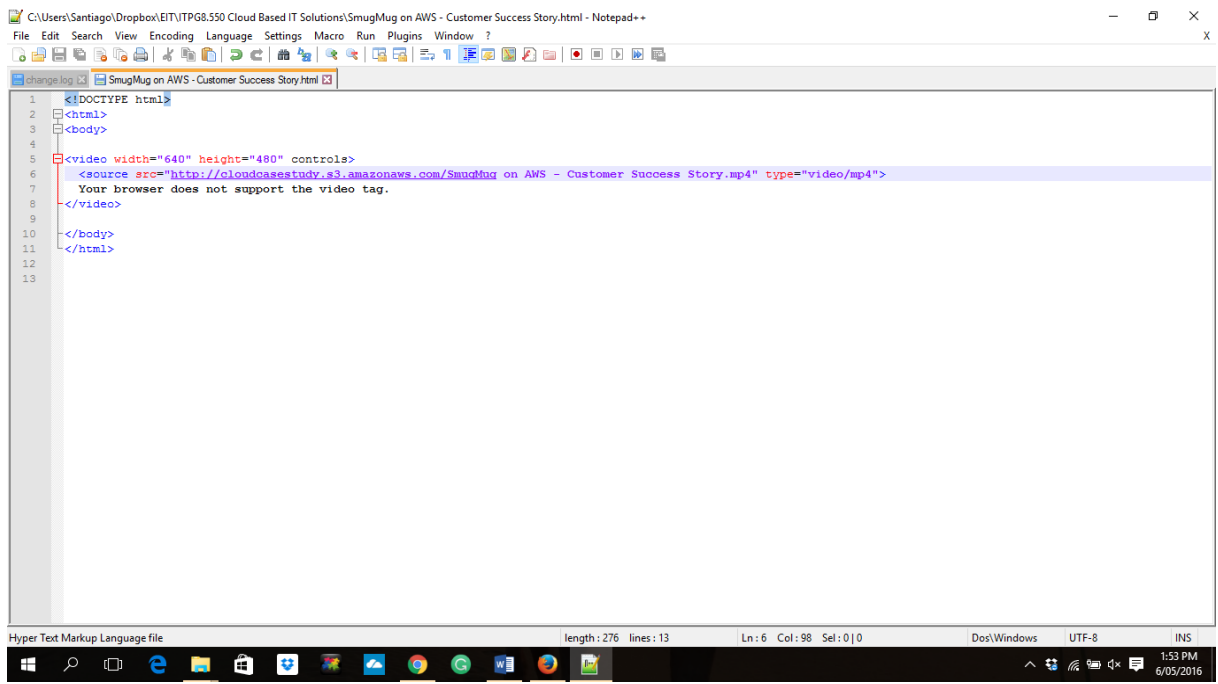

18. After the modification of the HTML file as shown above, the company website can link to the video hosted in AWS. However, the service is not because the file security options need to be changed. To modify the Access control List (ACL) option, please right click on the file and select the option "Edit ACL".

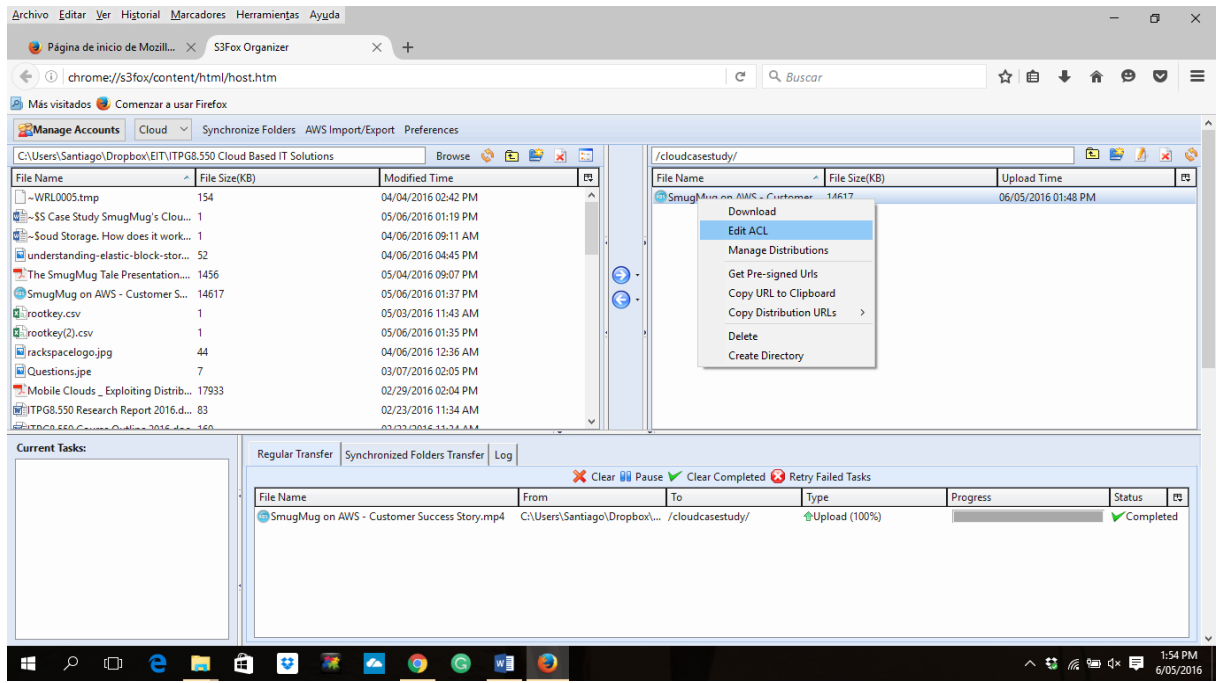

19. Change the settings to "Read" access for everyone, and press OK.

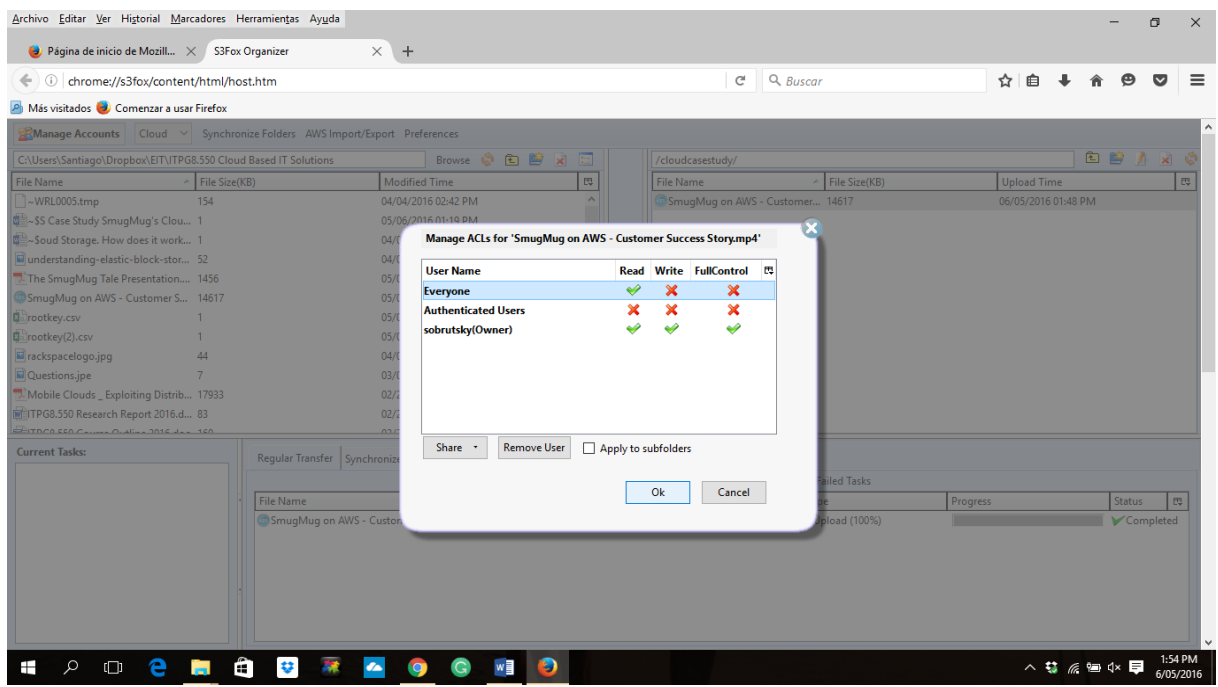

20. More videos can be added in the same manner, and can be managed using the S3Fox organizer extension.

**Summary and Further Considerations**

Amazon is the most popular provider of cloud infrastructure for businesses (Synergy Research Group, 2016), e.g. those that host customer data in the cloud. Amazon S3 provides security, reliability and high scalability. Amazon S3 also has a web service interface to store and retrieve data from the internet. One company which successfully implements Amazon S3 is SmugMug. It is a premium online photo and video sharing service business launched in 2002, which handles billions of photos and videos from amateur and professional photographers from all around the world. SmugMug uses AWS for its Amazon S3 storage capabilities as well as Amazon EC2, for better upload performance.

Amazon provides an overview of the AWS security policies and processes in a white paper, which is also available from Turning Technologies (2015). AWS cloud services are continuously audited by cloud service accreditation bodies. AWS is compliant with data protection laws in many countries, including New Zealand. AWS offers a "shared security model," whereby Amazon is responsible for protecting the underlying infrastructure and the user has the flexibility to choose and define the security controls for their resources (such as the encryption options, denial of service mitigation tools, and security monitoring features).

Therefore, AWS can be a good option for a mobile and online learning institute or a multimedia company, which either produces many videos or curates many videos from creative end users. Another important consideration while using cloud based storage for multimedia is around keeping videos for private use only for an intended audience, in order to protect copyrights and intellectual property (Erturk, 2013). Although YouTube is free and a popular platform, there are apparent limitations on how many private videos can be shared with users as well as requiring login to a Google account. This is where AWS can be configured to host and play an unlimited number of videos for many users within a company's own password secured website. Amazon S3 and EC2 also support a variety of video streaming formats, some of which are efficient for mobile device users. In conclusion, based on a preliminary literature review and hands-on testing, Amazon Cloud Storage appears to be a good backend solution for commercial companies and for online education providers.